\pdfoutput=1

\documentclass[11pt,twoside,a4paper,cmspaper,final,collab]{cms-tdr}
\def\svnVersion{33359:39664M}\def\cmsCernNo{2010-086}\def\cmsCernDate{\today}\def\cmsMessage{Submitted to Physical Review Letters}

\begin{document}\cmsNoteHeader{QCD-10-026}

\hyphenation{had-ron-i-za-tion}
\hyphenation{cal-or-i-me-ter}
\hyphenation{de-vices}
\RCS$Revision: 39635 $
\RCS$HeadURL: svn+ssh://alverson@svn.cern.ch/reps/tdr2/papers/QCD-10-026/trunk/QCD-10-026.tex $
\RCS$Id: QCD-10-026.tex 39635 2011-02-15 22:01:56Z apana $

\newcommand {\totlumi}{2.9~pb$^{-1}$\xspace}
\newcommand {\deltaphi}{$\Delta \varphi_\text{dijet}$\xspace}
\newcommand{\ptit}{\ensuremath{p_{T}}\xspace}
\newcommand{\rap}{\ensuremath{y}\xspace}
\newcommand {\ptmax}{\ensuremath{p_{T}}$^\text{max}$\xspace}
\newcommand {\pptmax}{\ensuremath{2p_{T}}$^\text{max}$\xspace}
\newcommand {\ptmaxx}{\ensuremath{p_{T}}$^\text{max}$$/2$\xspace}
\newcommand {\ptmaxns}{\ensuremath{p_{T}}$^\text{max}$}
\newcommand {\pptmaxns}{\ensuremath{2p_{T}}$^\text{max}$}
\newcommand {\ptmaxxns}{\ensuremath{p_{T}}$^\text{max}$$/2$}

\cmsNoteHeader{QCD-10-026} 
\title{\texorpdfstring{Dijet Azimuthal Decorrelations in pp Collisions at $\sqrt s$~=~7~TeV}{Dijet Azimuthal Decorrelations in pp Collisions at sqrt(s) = 7 TeV}}

\address[neu]{Northeastern University}
\address[fnal]{Fermilab}
\address[cern]{CERN}
\author[cern]{The CMS Collaboration}

\date{\today}

\abstract{Measurements of dijet azimuthal decorrelations in pp
collisions at $\sqrt s$ = 7~TeV using the CMS detector at the CERN
LHC are presented. The analysis is based on an inclusive dijet
event sample corresponding to an integrated luminosity of
2.9~pb$^{-1}$. The results are compared to predictions from
perturbative QCD calculations and various Monte Carlo event
generators. The dijet azimuthal distributions are found to be
sensitive to initial-state gluon radiation.}

\hypersetup{%
pdfauthor={CMS Collaboration},%
pdftitle={Dijet Azimuthal Decorrelations in pp Collisions at sqrt(s) = 7 TeV},%
pdfsubject={CMS},%
pdfkeywords={CMS, physics}}

\maketitle 

High-energy proton-proton collisions with high momentum transfer
are described within the framework of Quantum Chromodynamics (QCD)
as point-like scatterings between the proton constituents,
collectively referred to as partons. The outgoing partons manifest
themselves, through quark and gluon soft radiation and
hadronization processes, as localized streams of particles,
identified as jets. At Born level, dijets are produced with equal
transverse momenta \ptit with respect to the beam axis and
back-to-back in the azimuthal angle ($\Delta \varphi_\text{dijet}
= \left | \varphi_\text{jet1} - \varphi_\text{jet2} \right | =
\pi$). Soft-gluon emission will decorrelate the two highest \ptit
(leading) jets and cause small deviations from $\pi$. Larger
decorrelations from $\pi$ occur in the case of hard multijet
production.  Three-jet topologies dominate the region of
$2\pi/3<\Delta \varphi_\text{dijet} <\pi$, whereas angles smaller
than $2\pi/3$ are populated by four-jet events.

Dijet azimuthal decorrelations, \ie, the deviation of
$\Delta\varphi_\text{dijet}$ from $\pi$ for the two leading jets
in hard-scattering events can be used to study QCD radiation
effects over a wide range of jet multiplicities without the need
to measure all the additional jets. Such studies are important
because an accurate description of multiple-parton radiation is
still lacking in perturbative QCD (pQCD). Experiments therefore
rely on Monte Carlo (MC) event generators to take these
higher-order processes into account in searches for new physics
and for a wide variety of precision measurements. The observable
chosen to study the radiation effects is the differential dijet
cross section in $\Delta \varphi_\text{dijet}$, normalized by the
dijet cross section integrated over the entire $\Delta
\varphi_\text{dijet}$ phase space, $(1/ \sigma_\text{dijet})(d
\sigma_\text{dijet} / d \Delta \varphi_\text{dijet})$. By
normalizing the $\Delta \varphi_\text{dijet}$ distributions in
this manner, many experimental and theoretical uncertainties are
significantly reduced. Measurements of dijet azimuthal
decorrelations at the Tevatron have previously been reported by
the D0~collaboration~\cite{bib_d0-prl}. In this Letter, we present
the first measurements of dijet azimuthal decorrelations in pp
collisions at $\sqrt{s}=7$ TeV at the CERN Large Hadron Collider
(LHC).

The central feature of the Compact Muon Solenoid (CMS) apparatus
is a superconducting solenoid, of 6~m internal diameter, providing
an axial field of 3.8~T. Charged particle trajectories are
measured by the silicon pixel and strip tracker, covering
$0<\varphi<2\pi$ in azimuth and $|\eta| <$~2.5, where
pseudorapidity $\eta = -\text{ln [tan}(\theta/2)]$ and $\theta$ is
the polar angle relative to the counterclockwise proton beam
direction with respect to the center of the detector. A
lead-tungstate crystal electromagnetic calorimeter and a
brass/scintillator hadronic calorimeter surround the tracking
volume. The calorimeter cells are grouped in projective towers of
granularity $\Delta \eta \times \Delta \varphi = 0.087\times0.087$
at central pseudorapidities.  The granularity becomes coarser at
forward pseudorapidities. A preshower detector made of silicon
sensor planes and lead absorbers is installed in front of the
electromagnetic calorimeter at $1.653 <\vert \eta \vert < 2.6$.
Muons are measured in gas-ionization detectors embedded in the
steel magnetic field return yoke. A detailed description of the
CMS detector can be found elsewhere~\cite{bib_CMS}.

CMS uses a two-tiered trigger system to select events online;
Level-1 (L1) and the High Level Trigger (HLT). In this analysis,
events were selected using two inclusive single-jet triggers that
required an L1 jet with \ptit $>20$~GeV ($30$~GeV) and an HLT jet
with \ptit$>30$~GeV ($50$~GeV). The jets at L1 and HLT are
reconstructed using energies measured by the electromagnetic and
hadronic calorimeters and are not corrected for the jet energy
response of the calorimeters. The trigger efficiency for a given
corrected \ptit threshold of the leading jet (\ptmax) was measured
using events selected by a lower-threshold trigger. For the event
selection, \ptmax thresholds were chosen so that this efficiency
exceeded 99\%. The corresponding offline corrected \ptmax values
are 80~GeV (110~GeV) for the low (high) threshold jet trigger.

Jets were reconstructed offline using the anti-$k_T$ clustering
algorithm with a distance parameter $R = 0.5$~\cite{bib_antikt}.
The four-vectors of particles reconstructed by the CMS
particle-flow algorithm were used as input to the jet-clustering
algorithm. The particle-flow algorithm combines information from
all CMS sub-detectors to provide a complete list of long-lived
particles in the event. Muons, electrons, photons, and charged and
neutral hadrons are reconstructed individually. As a result, the
residual corrections to the jet four-vectors, arising from the
detector response, are relatively small (at the level of 5-10\% in
the central region)~\cite{bib_newjec}. A detailed description of
the particle-flow algorithm can be found
elsewhere~\cite{bib_pft09,bib_pft10}.

Spurious jets from noise and non-collision backgrounds were
eliminated by applying loose quality cuts on the jet
properties~\cite{bib_jetid}. Events were required to have a
primary vertex reconstructed along the beam axis and within 24 cm
of the detector center~\cite{bib_vertexpas}. Further cuts were
applied to reject interactions from the beam halo. Events were
selected having two leading jets each with \ptit $>30$~GeV and
rapidity $\left | \rap \right| < 1.1$, where $\rap = \frac{1}{2}
\text{ln} \left [ \left ( {E} + {p}_{z} \right)/ \left ( {E} -
{p}_{z} \right ) \right ]$, with $E$ being the total jet energy
and $p_{z}$ the projection of the jet momentum along the beam
axis. Each event is put into one of five mutually exclusive
regions, which are based on the \ptmax in the event.  The five
regions are: $80 <$ \ptmax $< 110$ GeV, $110 <$ \ptmax $< 140$
GeV, $140 <$ \ptmax $< 200$ GeV, $200 <$ \ptmax $< 300$ GeV, and
$300$ GeV $<$ \ptmax. The data correspond to an integrated
luminosity of 0.3~pb$^{-1}$ for the lowest \ptmax region and
2.9~pb$^{-1}$ for the other \ptmax regions.  The uncertainty on
the integrated luminosity is estimated to be 11\%~\cite{bib_lumi}.
After the application of all selection criteria, the numbers of
events remaining in each of the five \ptmax regions, starting from
the lowest, are: $60837$, $160388$, $69009$, $14383$, and $2284$.

The $\Delta \varphi_\text{dijet}$ distributions are corrected for
event migration effects due to the finite jet \ptit and position
resolutions of the detector. The distributions are sensitive to
the jet \ptit resolution because fluctuations in the jet response
can cause low-energy jets to be misidentified as leading jets, and
events can migrate between different \ptmax regions. The finite
resolution in azimuthal angle causes event migration between
$\Delta \varphi_\text{dijet}$ bins, while the resolution in
rapidity can move jets in and out of the central rapidity region
($\left | \rap \right | < 1.1$). The correction factors were
determined using two independent MC samples: {\sc pythia} 6.422
({\sc pythia6})~\cite{bib_pythia} tune {\sc
D6T}~\cite{bib_pythiatunes}, and {\sc herwig++}
2.4.2~\cite{bib_herwig}. The \ptit, rapidity, and azimuthal angle
of each generated jet were smeared according to the measured
resolutions~\cite{bib_jec}.  The ratio of the two dijet azimuthal
distributions (the generated distribution and the smeared one)
determined the unfolding correction factors for each \ptmax
region, for a given MC sample. The average of the correction
factors for each \ptmax region from the two MC samples were used
as the final unfolding correction applied to data. The unfolding
correction factors modify the measured $\Delta
\varphi_\text{dijet}$ distributions by less than 2\% for $5 \pi /6
<\Delta \varphi_\text{dijet}<\pi$.  For $\Delta
\varphi_\text{dijet} \sim \pi/2$, the changes range from $-11$\%,
for the highest \ptmax region, to $-19\%$, for the lowest.

The main sources of systematic uncertainty arise from
uncertainties in the jet energy calibration, the jet \ptit
resolution, and the unfolding correction. The jet energy
calibration uncertainties have been tabulated for the considered
phase space in the variables of jet \ptit and
$\eta$~\cite{bib_newjec}. Typical values are between 2.5\% and
3.5\%. The resulting uncertainties on the normalized
$\Delta\varphi_\text{dijet}$ distributions range from 5\% at
$\Delta \varphi_\text{dijet} \sim \pi/2$ to 1\% at $\Delta
\varphi_\text{dijet} \sim \pi$.  The effect of jet \ptit
resolution uncertainty on the $\Delta \varphi_\text{dijet}$
distributions was estimated by varying the jet \ptit resolutions
by $\pm 10$\%~\cite{bib_jec} and comparing the $\Delta
\varphi_\text{dijet}$ unfolding correction before and after the
change. This yields a variation on the normalized $\Delta
\varphi_\text{dijet}$ distributions ranging from 5\% at $\Delta
\varphi_\text{dijet} \sim \pi/2$ to 1\% at $\Delta
\varphi_\text{dijet} \sim \pi$.  The uncertainties on the
unfolding correction factors were estimated by comparing the
corrections from different event generators and {\sc pythia6}
tunes that vary significantly in their modelling of the jet
kinematic distributions and $\Delta \varphi_\text{dijet}$
distributions. The resulting uncertainty varies from 8\% at
$\Delta \varphi_\text{dijet} \sim \pi/2$ to 1.5\% at $\Delta
\varphi_\text{dijet} \sim \pi$. The systematic uncertainty from
using a parametrized model to simulate the finite jet \ptit and
position resolutions of the detector to determine the unfolding
correction factors was estimated to be about 2.5\% in all \ptmax
regions. The combined systematic uncertainty, calculated as the
quadratic sum of all systematic uncertainties, varies from 11\% at
$\Delta \varphi_\text{dijet} \sim \pi/2$ to 3\% at $\Delta
\varphi_\text{dijet} \sim \pi$.

\begin{figure}[htb]
    \centering
    \includegraphics[width=\linewidth]{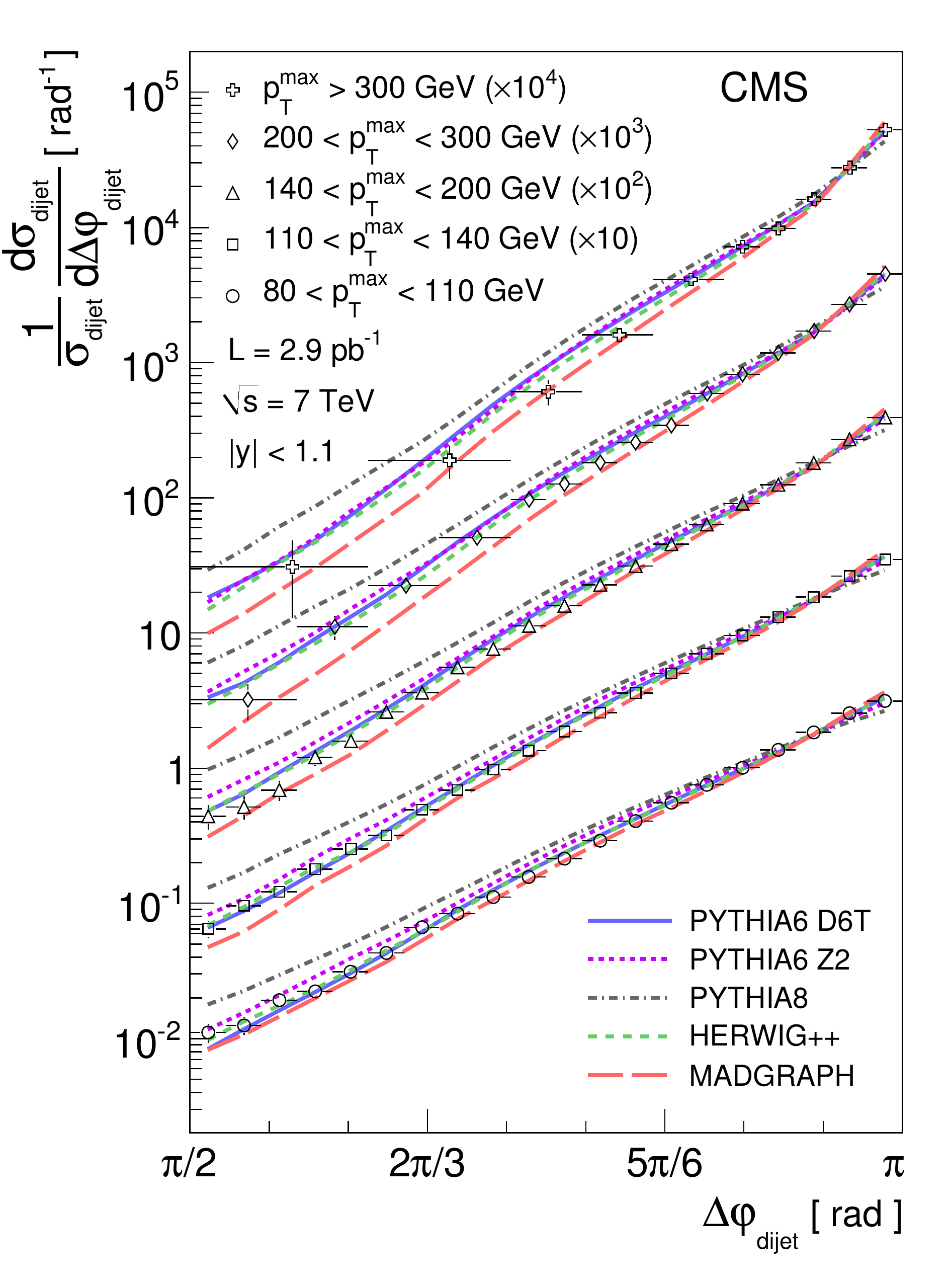}
    \caption{
        Normalized \deltaphi distributions in several \ptmax regions, scaled by the multiplicative
        factors given in the figure for easier presentation. The curves
        represent predictions from {\sc pythia6}, {\sc pythia8}, {\sc herwig++}, and {\sc madgraph}.
        The error bars on the data points include statistical and systematic
        uncertainties.}
    \label{fig_dPhi}
\end{figure}

\begin{figure}[htb]
    \centering
    \includegraphics[width=\linewidth]{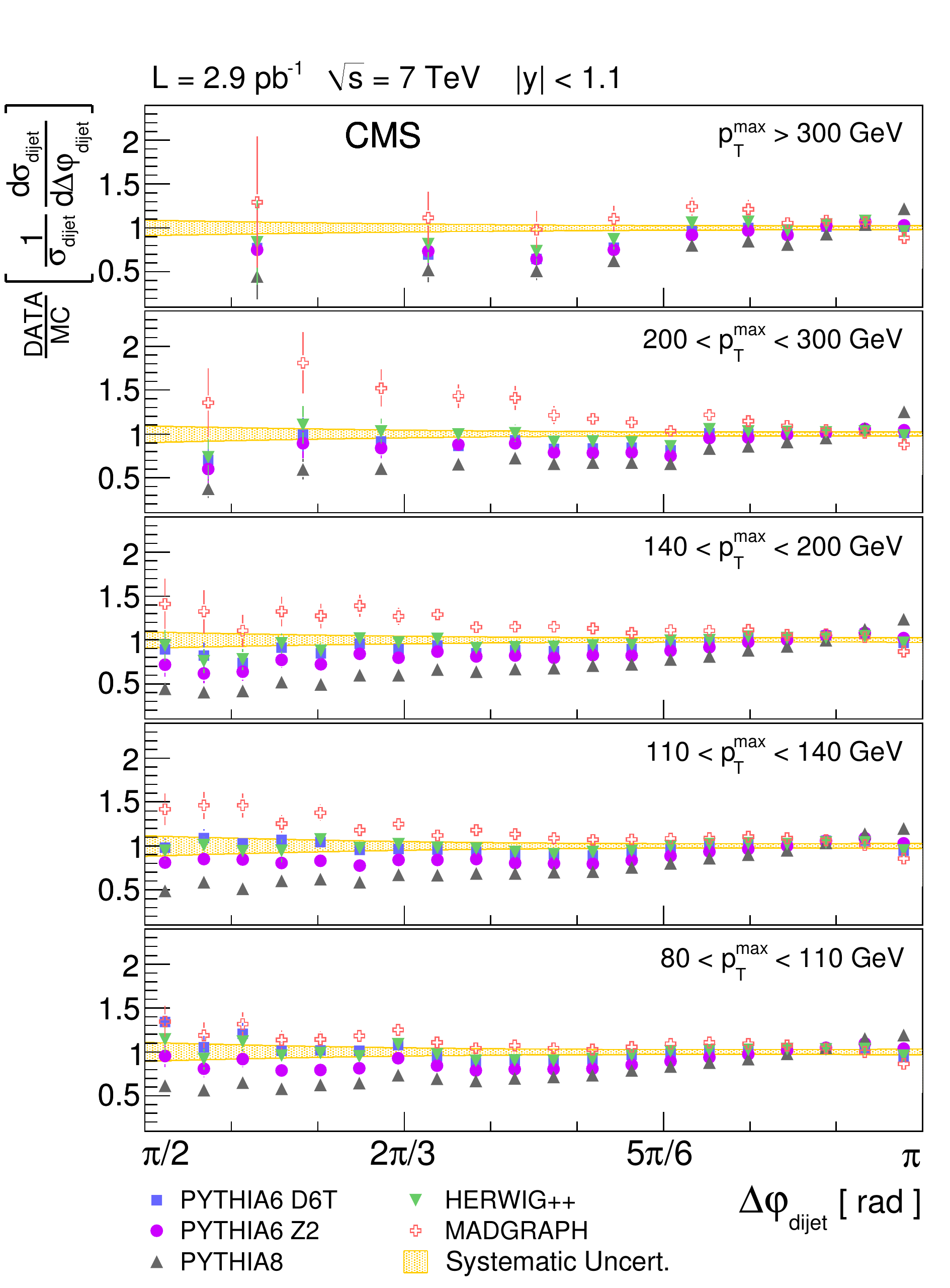}
    \caption{Ratios of measured normalized \deltaphi distributions to
        {\sc pythia6}, {\sc pythia8}, {\sc herwig++}, and {\sc madgraph} predictions in several \ptmax
        regions. The shaded bands indicate the total systematic
uncertainty.
    }
    \label{fig_dPhi_ratio}
\end{figure}

\ifthenelse{\boolean{cms@external}}{}
{
\begin{figure}[htb]
    \centering
    \includegraphics[width=\linewidth]{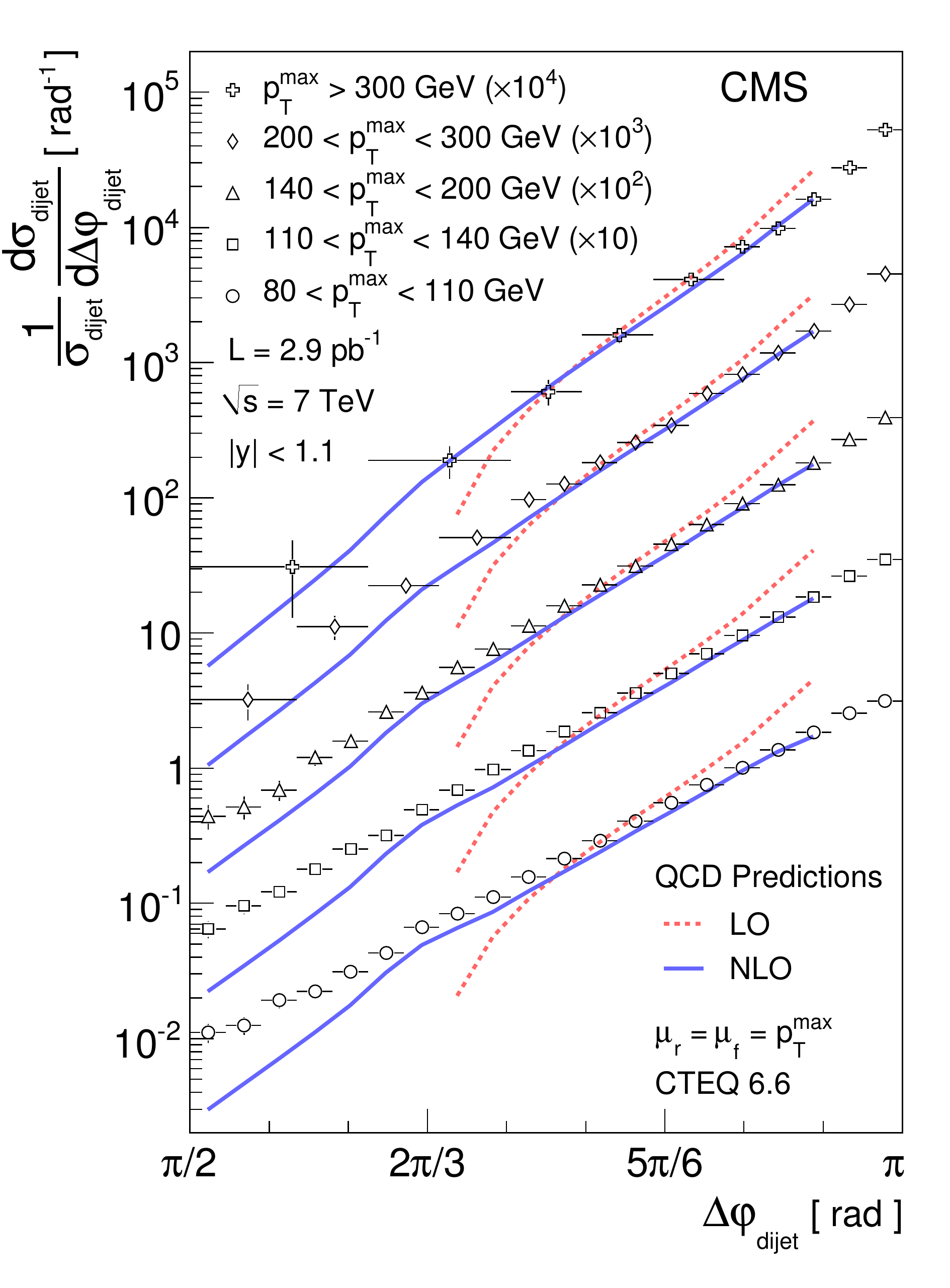}
    \caption{
        Normalized \deltaphi distributions in several \ptmax regions, scaled by the multiplicative
        factors given in the figure for easier presentation. The curves
        represent predictions from LO (dotted line) and NLO pQCD (solid line).  Non-perturbative corrections
        have been applied to the predictions. The
        error bars on the data points include statistical and systematic
        uncertainties.}
    \label{fig_dPhi_NLO}
\end{figure}
}
\begin{figure}[htb]
    \centering
    \includegraphics[width=\linewidth]{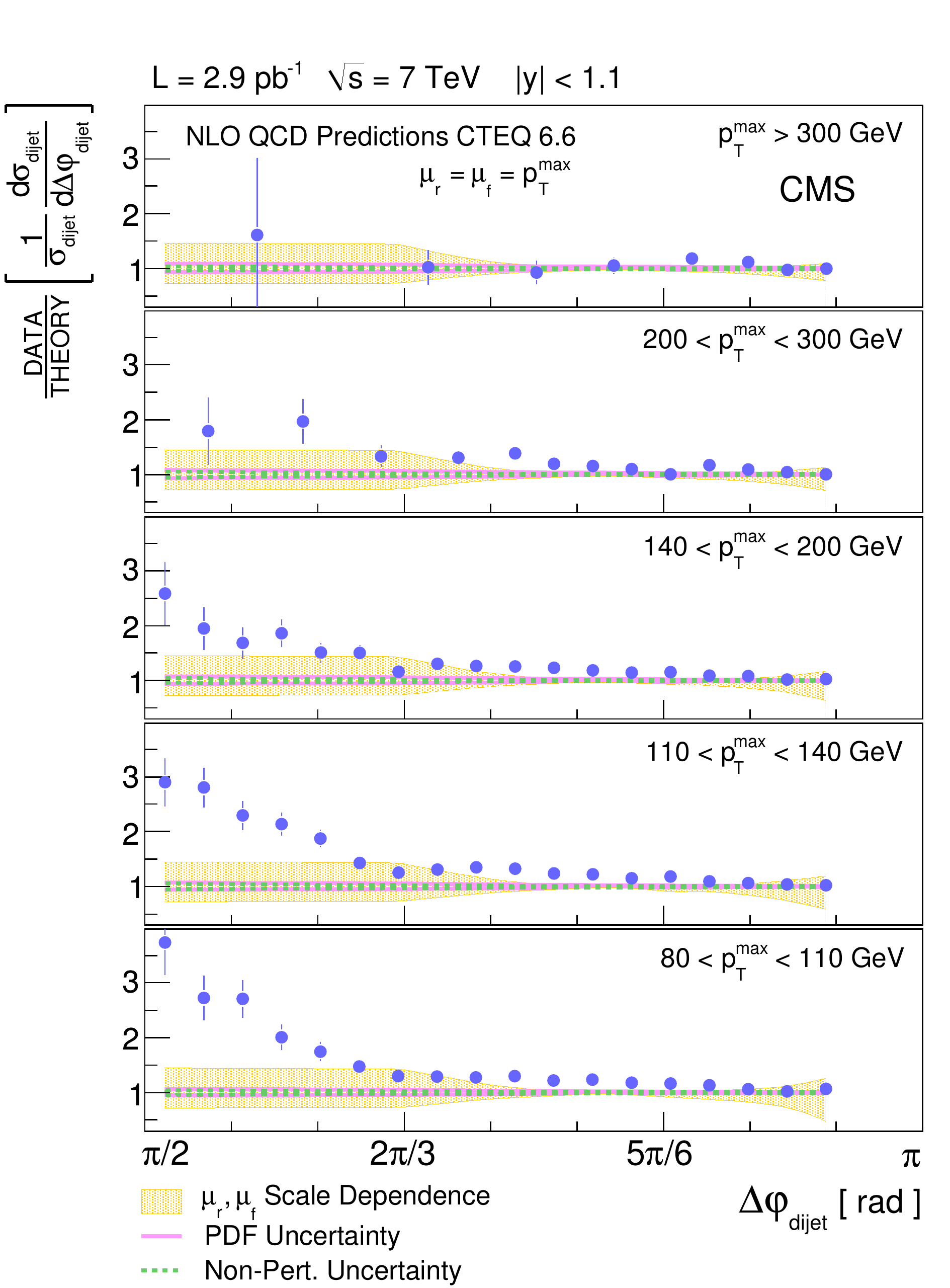}
    \caption{Ratios of measured normalized \deltaphi distributions to
        NLO pQCD predictions with non-perturbative corrections in several
        \ptmax regions. The error bars on the data points include statistical and
        systematic uncertainties.  The effect on the NLO pQCD predictions due to $\mu_r$ and $\mu_f$ scale variations and
        PDF uncertainties,
        as well as the uncertainties from the non-perturbative corrections are shown.
    }
    \label{fig_dPhi_ratio_NLO}
\end{figure}

\begin{figure}[htb]
    \centering
    \includegraphics[width=\linewidth]{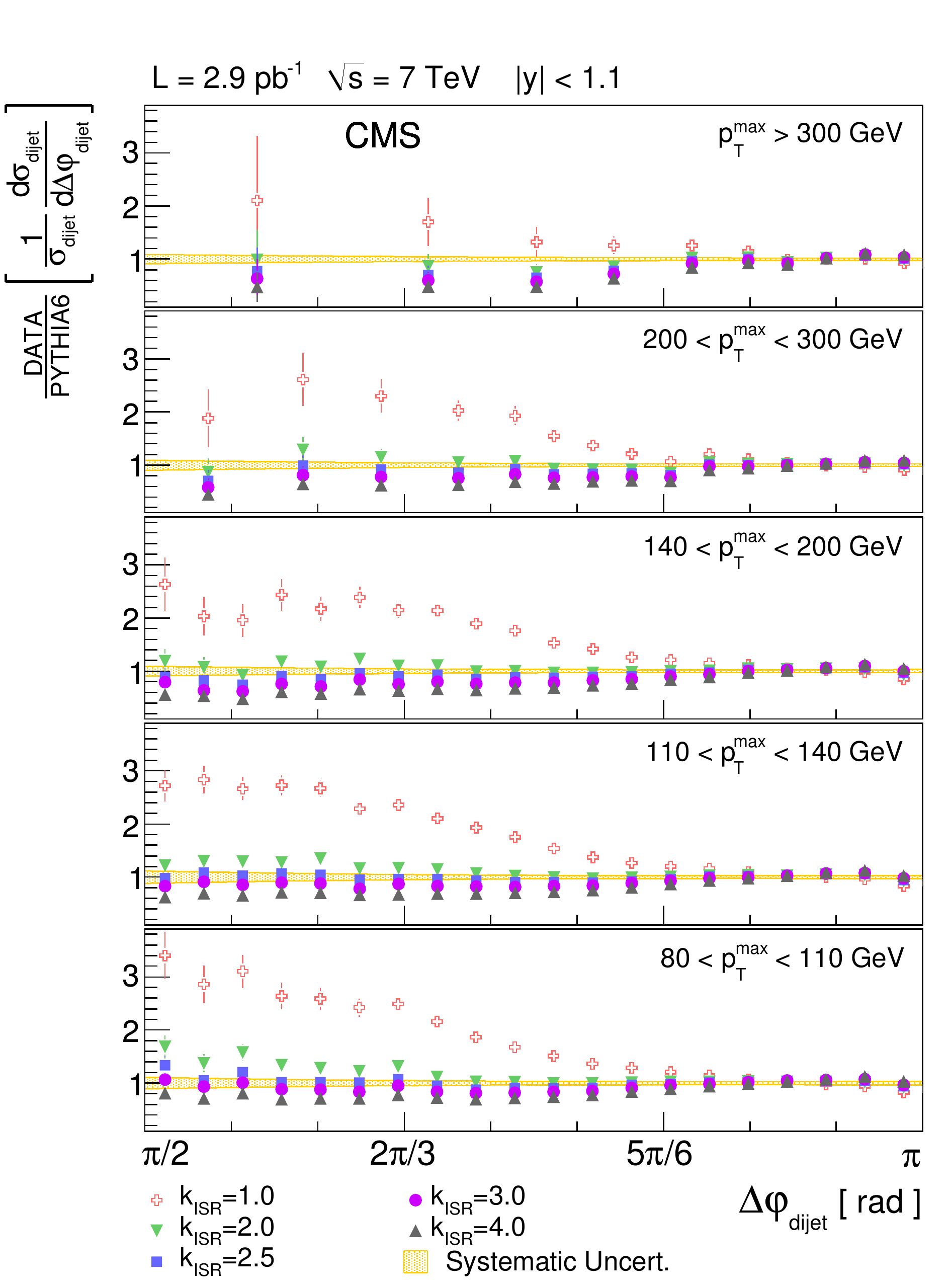}
    \caption{Ratios of measured normalized \deltaphi distributions to {\sc pythia6} tune
     {\sc D6T} with various values of $k_{\text{ISR}}$ in several \ptmax regions. The shaded bands
     indicate the total systematic uncertainty.}
    \label{fig_dPhi_ISR}
\end{figure}

The corrected differential \deltaphi distributions, normalized to
the integrated dijet cross section, are shown in
Fig.~\ref{fig_dPhi} for the five \ptmax regions. The distributions
are scaled by multiplicative factors for presentation purposes.
Each data point is plotted at the abscissa value for which the
predicted differential $\Delta \varphi_\text{dijet}$ distribution
has the same value as the bin average obtained using {\sc pythia6}
tune {\sc D6T}, which provides a good description of the
data~\cite{bib_bin}.

The \deltaphi distributions are strongly peaked at $\pi$ and
become steeper with increasing \ptmax. The simulated \deltaphi
distributions from the {\sc pythia6} ({\sc D6T} and {\sc
Z2}~\cite{bib_tunez2} tunes), {\sc pythia 8.135} ({\sc
pythia8})~\cite{bib_pythia8}, {\sc herwig++}, and {\sc madgraph
4.4.32}~\cite{bib_madgraph} event generators are presented for
comparison. The {\sc madgraph} generator is based on leading-order
matrix element multiparton final-state predictions, using {\sc
pythia6} for parton showering and hadronization, and the MLM
method~\cite{bib_mlm} to map the parton-level event into a parton
shower history.  The {\sc madgraph} predictions included
tree-level processes of up to four partons. For {\sc pythia6},
{\sc pythia8}, and {\sc madgraph} event generators the {\sc
CTEQ6L} ~\cite{bib_cteq} parton distribution functions (PDFs) were
used; for {\sc herwig++}, the {\sc MRST2001} PDFs~\cite{bib_mrst}.

Figure~\ref{fig_dPhi_ratio} shows the ratios of the measured
\deltaphi distributions to the predictions of {\sc pythia6}, {\sc
pythia8}, {\sc herwig++}, and {\sc madgraph} in the five \ptmax
regions. The combined systematic uncertainty on the experimental
measurements is shown by the shaded band. The predictions from
{\sc pythia6} and {\sc herwig++} describe the shape of the data
distributions well, while {\sc madgraph} ({\sc pythia8}) predicts
less (more) azimuthal decorrelation than is observed in the data.

\ifthenelse{\boolean{cms@external}}
{
Figure~\ref{fig_dPhi_ratio_NLO} displays the ratios of the
measured \deltaphi distributions to the next-to-leading-order
(NLO) predictions of pQCD calculations from the parton-level
generator {\sc nlojet++}~\cite{bib_nlojet} within the {\sc
fastnlo} framework~\cite{bib_fastnlojet}. The predictions include
$2 \rightarrow 3$ processes at NLO, normalized to
$\sigma_\text{dijet}$ at NLO\@:

\begin{equation*}
  \frac{1}{\sigma_\text{dijet}} \left| \frac{}{}_\text{NLO} \times
    \frac{d \sigma_\text{dijet}}{d \Delta \varphi_\text{dijet}} \right|_\text{NLO}.
\end{equation*}
The predictions near $\Delta \varphi_\text{dijet} = \pi$ have been
excluded because of their sensitivity to higher-order corrections
not included in the present calculations.
}
{Figure~\ref{fig_dPhi_NLO} displays a comparison between the
measured \deltaphi distributions and the predictions of pQCD
calculations from the parton-level generator {\sc
nlojet++}~\cite{bib_nlojet} within the {\sc fastnlo}
framework~\cite{bib_fastnlojet}. The predictions near $\Delta
\varphi_\text{dijet} = \pi$ have been excluded because of their
sensitivity to higher-order corrections not included in the
present calculations. The leading-order (LO) curves represent
processes with three partons in the final state, normalized to the
LO $\sigma_\text{dijet}$ ($2 \rightarrow 2$) cross section. The
next-to-leading-order (NLO) predictions include $2 \rightarrow 3$
processes at NLO, normalized to $\sigma_\text{dijet}$ at NLO\@:

\begin{equation*}
  \frac{1}{\sigma_\text{dijet}} \left| \frac{}{}_\text{(N)LO} \times
    \frac{d \sigma_\text{dijet}}{d \Delta \varphi_\text{dijet}} \right|_\text{(N)LO}.
\end{equation*}

}

Uncertainties due to the renormalization ($\mu_{r}$) and
factorization ($\mu_{f}$) scales are evaluated by varying the
default choice of $\mu_{r} = \mu_{f}$ = \ptmax between \ptmaxx and
\pptmax in the following six combinations: $(\mu_{r},\mu_{f}) =
($\ptmaxxns, \ptmaxxns$)$, $($\pptmaxns, \pptmaxns$)$,
$($\ptmaxns, \ptmaxxns$)$, $($\ptmaxns, \pptmaxns$)$,
$($\ptmaxxns, \ptmaxns$)$, and $($\pptmaxns, \ptmaxns$)$. These
scale variations modify the predictions of the normalized $\Delta
\varphi_\text{dijet}$ distributions by less than 50\%. The PDFs
and the associated uncertainties were obtained from {\sc
CTEQ6.6}~\cite{bib_cteq}.  The PDF uncertainties were derived
using the 22 {\sc CTEQ6.6} uncertainty eigenvectors and found to
be 9\% at $\Delta \varphi_\text{dijet} \sim \pi/2$ and 2\% at
$\Delta \varphi_\text{dijet} < \pi$. Following the proposal of the
PDF4LHC working group~\cite{Alekhin:2011sk}, the impact of other
global PDF fits~\cite{Martin:2009iq,Ball:2010de,Lai:2010vv} was
investigated and found to be negligible in the context of this
analysis.

Non-perturbative corrections due to hadronization and
multiple-parton interactions were applied to the pQCD predictions.
The correction factors were determined from the {\sc pythia6} and
{\sc herwig++} simulations and modify the predictions from +4\%
($\Delta \varphi_\text{dijet} \sim \pi$) to $-13$\% ($\Delta
\varphi_\text{dijet} \sim \pi/2$).  The uncertainty due to the
non-perturbative corrections is estimated to be 6\% at $\Delta
\varphi_\text{dijet} \sim \pi/2$ and 2\% at $\Delta
\varphi_\text{dijet} \sim \pi$.\ifthenelse{\boolean{cms@external}}
{The effect due to the scale variations, as well as the uncertainties
due to PDFs and non-perturbative corrections, are also shown in
Fig.~\ref{fig_dPhi_ratio_NLO}.}
{The ratios of the measured
\deltaphi distributions to the NLO pQCD predictions are shown in
Fig.~\ref{fig_dPhi_ratio_NLO}. The effect due to the scale
variations, as well as the uncertainties due to PDFs and
non-perturbative corrections, are also shown.}
The NLO predictions
provide a good description of the shape of the data distributions
over much of the \deltaphi range. Compared to the data, the
reduced decorrelation in the theoretical prediction and the
increased sensitivity to the $\mu_r$ and $\mu_f$ scale variations
for $\Delta \varphi_\text{dijet} < 2\pi/3$ shown in
Fig.~\ref{fig_dPhi_ratio_NLO} are attributed to the fact that the
pQCD prediction in this region is effectively available only at
leading order, since the contribution from tree-level four-parton
final states dominates.

The sensitivity of the \deltaphi distributions to initial-state
parton shower radiation (ISR) is investigated by varying the input
parameter $k_{\text{ISR}}$ ({\sc parp(67)}) in {\sc pythia6} tune
{\sc D6T}. The product of $k_{\text{ISR}}$ and the square of the
hard-scattering scale gives the maximum allowed parton virtuality
(i.e., the maximum allowed \ptit) in the initial-state shower.
Previous studies have shown that $k_{\text{ISR}}$ is the only
parameter in {\sc pythia6} that has significant impact on the
\deltaphi distributions~\cite{bib_LHC-report}. The default value
of $k_{\text{ISR}}$ in {\sc pythia6} tune {\sc D6T} is 2.5,
determined from the D0 dijet azimuthal decorrelation
results~\cite{bib_d0-prl}. Figure~\ref{fig_dPhi_ISR} shows
comparisons of the measured \deltaphi distributions to {\sc
pythia6} distributions with various $k_{\text{ISR}}$ values. The
effects are more pronounced for smaller \deltaphi angles, where
multi-gluon radiation dominates.  Varying $k_{\text{ISR}}$ by $\pm
0.5$ about its default value yields a change of about 30\% on the
{\sc pythia6} prediction for $\Delta \varphi_\text{dijet} \sim
\pi/2$, suggesting that our results could be used to tune
parameters in the MC event generators that control radiative
effects in the initial state.  In {\sc pythia6} tune {\sc D6T},
the maximum \ptit allowed in final-state radiation parton shower
is controlled through the parameter {\sc parp(71)}. We varied the
value of this parameter from 2.5 to 8 (the default value is 4.0)
and observed less than $\sim 10$\% changes in the $\Delta
\varphi_\text{dijet}$ distributions in all \ptit regions.

In summary, we have measured dijet azimuthal decorrelations in
different leading-jet \ptit regions from pp collisions at $\sqrt
s$ = 7~TeV. The {\sc pythia6} and {\sc herwig++} event generators
are found to best describe the shape of the measured distributions
over the entire range of \deltaphi. The predictions from NLO pQCD
are in reasonable agreement with the measured distributions,
except at small \deltaphi where multiparton radiation effects
dominate. The \deltaphi distributions are found to be sensitive to
initial-state gluon radiation.

We wish to congratulate our colleagues in the CERN accelerator
departments for the excellent performance of the LHC machine. We
thank the technical and administrative staff at CERN and other CMS
institutes, and acknowledge support from: FMSR (Austria); FNRS and
FWO (Belgium); CNPq, CAPES, FAPERJ, and FAPESP (Brazil); MES
(Bulgaria); CERN; CAS, MoST, and NSFC (China); COLCIENCIAS
(Colombia); MSES (Croatia); RPF (Cyprus); Academy of Sciences and
NICPB (Estonia); Academy of Finland, ME, and HIP (Finland); CEA
and CNRS/IN2P3 (France); BMBF, DFG, and HGF (Germany); GSRT
(Greece); OTKA and NKTH (Hungary); DAE and DST (India); IPM
(Iran); SFI (Ireland); INFN (Italy); NRF and WCU (Korea); LAS
(Lithuania); CINVESTAV, CONACYT, SEP, and UASLP-FAI (Mexico); PAEC
(Pakistan); SCSR (Poland); FCT (Portugal); JINR (Armenia, Belarus,
Georgia, Ukraine, Uzbekistan); MST and MAE (Russia); MSTD
(Serbia); MICINN and CPAN (Spain); Swiss Funding Agencies
(Switzerland); NSC (Taipei); TUBITAK and TAEK (Turkey); STFC
(United Kingdom); DOE and NSF (USA).

\bibliography{auto_generated}   

\begin{thebibliography}{10}
\bibitem{bib_tunez2}
The {\sc pythia6} {\sc Z2} tune is identical to the {\sc Z1} tune described
  in~\cite{bib_pythiatunes} except that {\sc Z2} uses the CTEQ6L PDF while {\sc
  Z1} uses CTEQ5L.
\end{thebibliography}
\cleardoublepage\appendix\section{The CMS Collaboration \label{app:collab}}\begin{sloppypar}\hyphenpenalty=5000\widowpenalty=500\clubpenalty=5000\textbf{Yerevan Physics Institute,  Yerevan,  Armenia}\\*[0pt]
V.~Khachatryan, A.M.~Sirunyan, A.~Tumasyan
\vskip\cmsinstskip
\textbf{Institut f\"{u}r Hochenergiephysik der OeAW,  Wien,  Austria}\\*[0pt]
W.~Adam, T.~Bergauer, M.~Dragicevic, J.~Er\"{o}, C.~Fabjan, M.~Friedl, R.~Fr\"{u}hwirth, V.M.~Ghete, J.~Hammer\cmsAuthorMark{1}, S.~H\"{a}nsel, C.~Hartl, M.~Hoch, N.~H\"{o}rmann, J.~Hrubec, M.~Jeitler, G.~Kasieczka, W.~Kiesenhofer, M.~Krammer, D.~Liko, I.~Mikulec, M.~Pernicka, H.~Rohringer, R.~Sch\"{o}fbeck, J.~Strauss, A.~Taurok, F.~Teischinger, P.~Wagner, W.~Waltenberger, G.~Walzel, E.~Widl, C.-E.~Wulz
% ...
\vskip\cmsinstskip
% ...
1:~~Also at CERN, European Organization for Nuclear Research, Geneva, Switzerland\\
% ...

\end{sloppypar}
\end{document}